\documentclass[aps,prc,twocolumn,10pt, superscriptaddress, showpacs, floatfix]{revtex4-1}
\usepackage{amssymb,epsfig}
\usepackage{xcolor}
\graphicspath{{figures/}}
\newcommand{\be}{\begin{equation}}
\newcommand{\ee}{\end{equation}}
\newcommand{\bea}{\begin{eqnarray}}
\newcommand{\eea}{\end{eqnarray}}

\begin{document}

\title{Temperature dependence of nuclear spin-isospin response and beta decay in hot astrophysical environments
}
\author{Elena Litvinova}
\affiliation{Department of Physics, Western Michigan University, Kalamazoo, MI 49008, USA}
\affiliation{National Superconducting Cyclotron Laboratory, Michigan State University, East Lansing, MI 48824, USA}
\author{Caroline Robin}
\affiliation{Institute for Nuclear Theory, University of Washington, Seattle, WA 98195, USA }
\affiliation{JINA-CEE, Michigan State University, East Lansing, MI 48824, USA}
\author{Herlik Wibowo}
\affiliation{Department of Physics, Western Michigan University, Kalamazoo, MI 49008, USA}

\date{\today}

\begin{abstract}
A microscopic approach to the proton-neutron nuclear response is formulated in the finite-temperature relativistic nuclear field theory framework. The approach is based on the meson-nucleon Lagrangian of quantum hadrodynamics and advances the relativistic field theory  for spin-isospin response beyond the finite-temperature random phase approximation. The dynamical contribution to the in-medium proton-neutron interaction amplitude is described in a parameter-free way by the coupling between the single nucleons and 
strongly-correlated particle-hole excitations (phonons) within the newly developed finite-temperature formalism. In this framework we investigate temperature dependence of the Gamow-Teller and spin dipole resonances in the closed-shell nuclei  $^{48}$Ca, $^{78}$Ni, and $^{132}$Sn. Broader impacts of their temperature dependence are illustrated for the associated beta decay rates and lifetimes of $^{78}$Ni and $^{132}$Sn in hot astrophysical environments. We 
found a remarkable sensitivity of the beta decay rates to the enhanced low-energy spin-isospin strength at finite temperature, in particular, to the contribution of the first-forbidden transitions.

\end{abstract}

\maketitle

{\it Introduction. \textemdash} 
Response to charge-changing, or isospin-flip, probes, which induce a conversion of a nucleon of one type to another (proton to neutron or neutron to proton), is one of the most important characteristics of nuclear systems. On the fundamental level this type of response provides information on nuclear weak interactions and underlying forces in the proton-neutron channel, and in the context of applications it has a very broad impact on nuclear sciences from nuclear data \cite{Brown1985,Langanke2001,iaea} to astrophysics \cite{Thielemann2001,MumpowerSurmanMcLaughlinEtAl2016}.

Astrophysical implications of the isospin-transfer excitations include, for instance, beta decay, electron capture, neutrino capture and scattering, which occur under different conditions formed in various stages of star evolution and merging of neutron stars.
The cross sections and rates of these processes within a broad range of densities and temperatures are decisive for astrophysical modeling \cite{Cowan1991b,Arnould2007,MumpowerSurmanMcLaughlinEtAl2016}. Some of them can be determined in laboratory experiments, however, many exotic nuclear systems  
located far away from the beta-stability valley of the nuclear chart are beyond the present and even future experimental capabilities.
Therefore, reliable theoretical predictions are needed for the isospin-transfer excitations, such as the Gamow-Teller response (GTR) with transfer of one unit of isospin and one unit of spin, the spin-dipole response with transfer of an additional unit of angular momentum, and the pure isospin-flip isobaric-analog resonance at finite temperatures.   

A theoretical description of these processes can be provided, for instance, by the shell-model or by the shell-model Monte-Carlo approach combined with the random phase approximation (RPA) \cite{Langanke2001,Langanke2003,Pinedo2014}.  The predictive shell-model calculations are, however, very difficult to be extended beyond the pf-shell. The RPA, in turn, is very limited in the treatment of many-body correlations. Theoretical approaches to the proton-neutron nuclear response at finite-temperature are mostly confined by the finite-temperature quasiparticle RPA (FT-QRPA) \cite{Minato2009,Dzhioev2010} or the finite-temperature relativistic random phase approximation (FT-RRPA) \cite{Niu2011} which provide a convenient framework for studying the Gamow-Teller (GT) and first-forbidden (FF) strength distributions in both (p,n) and (n,p) channels, beta-decay rates \cite{Minato2009}, and electron capture rates \cite{Dzhioev2010,Niu2011}. Pairing correlations taken into account in the FT-QRPA are important for the temperatures below the critical temperature which typically amounts to 0.5-1.0 MeV in medium-heavy nuclei. However, the (Q)RPA theories are, in principle, limited by the one-fermion loop approximation and can not account for important retardation effects which are responsible for the damping effects. At zero temperature, they are sometimes solely responsible for the decay of neutron-rich nuclei and necessary for accurate predictions of weak nuclear processes in fully self-consistent theories \cite{MarketinLitvinovaVretenarEtAl2012,NiuNiuColoEtAl2015,RobinLitvinova2016,Niu2018}. 
%

In order to meet the very high standards required for nuclear science applications, theoretical approaches to the nuclear response must include correlations beyond (Q)RPA and, at the same time, be based on fundamental concepts of the nucleon-nucleon interaction. The latter provides an advanced predictive power and the former is of the utmost importance as the inaccuracies contained in nuclear strength functions can propagate tremendously \cite{MumpowerSurmanMcLaughlinEtAl2016}. 
In this Letter, we present a novel approach to the finite-temperature proton-neutron nuclear response, which is going towards these requirements.  We advance the approach developed previously in the zero-temperature framework of the relativistic nuclear field theory (RNFT) \cite{LitvinovaRing2006, LitvinovaRingTselyaev2007,LitvinovaRingTselyaev2008,LitvinovaRingTselyaev2010,MarketinLitvinovaVretenarEtAl2012,LitvinovaBrownFangEtAl2014,RobinLitvinova2016,Litvinova2016} to the finite-temperature case. The RNFT is based on the covariant energy density functional \cite{VretenarAfanasjevLalazissisEtAl2005} and extends both the neutral-channel and proton-neutron relativistic RPA (pn-RRPA) \cite{RingMaVanGiaiEtAl2001,Kurasawa2003} beyond the one-loop approximation by taking into account the medium polarization effects in the form of the particle-vibration coupling (PVC) in a parameter-free way. It was found that these effects play a crucial role in describing the nuclear response in both neutral 
\cite{LitvinovaRingTselyaevEtAl2009,LitvinovaLoensLangankeEtAl2009,LitvinovaRingTselyaev2010,EndresLitvinovaSavranEtAl2010,
TamiiPoltoratskaNeumannEtAl2011,MassarczykSchwengnerDoenauEtAl2012,LitvinovaRingTselyaev2013,SavranAumannZilges2013,LitvinovaRingTselyaev2013,LanzaVitturiLitvinovaEtAl2014,PoltoratskaFearickKrumbholzEtAl2014,Oezel-TashenovEndersLenskeEtAl2014,EgorovaLitvinova2016} and charge-exchange \cite{MarketinLitvinovaVretenarEtAl2012,LitvinovaBrownFangEtAl2014,RobinLitvinova2016,Litvinova2018} channels. Recently, the theory for neutral excitations was extended to finite temperatures \cite{LitvinovaWibowo2018}, and here we present another extension for the thermal proton-neutron response.

{\it Method. \textemdash} 
The finite-temperature relativistic mean-field (RMF)
theory based on the minimization of the grand potential $\Omega(\mu,T)$ \cite{Sommermann1983} 
\be
\Omega(\mu,T) = E  - T S - \mu N
\label{Omega}
\ee
is applied to calculate microscopic characteristics of the initial compound nucleus at finite temperature. 
The grand potential is minimized with the Lagrange multipliers $\mu$ and $T$ determined by the
average energy $E$, particle number $N$, and the entropy $S$. The latter two
quantities are thermal averages with the 
one-body nucleonic density operator $\hat{\rho}$ of trace unity:
\be
S = -k\text{Tr}({\hat{\rho}}\text{ln}\hat{\rho}), \ \ \ 
\ N= \text{Tr}(\hat{\rho} {\hat{\cal N}}),
\label{SN}
\ee 
where $\hat{\cal N}$ is the particle number operator, and $k$ is the Boltzmann constant which is equal to one in the natural units.
The energy is a covariant functional of the nucleonic density and classical meson and photon fields $\phi_m$ \cite{VretenarAfanasjevLalazissisEtAl2005}:
\bea
E[\hat{\rho},\phi_m] &=& \text{Tr}[({\vec\alpha}\cdot{\vec p} + \beta M)\hat{\rho}] + \sum\limits_m\Bigl\{\text{Tr}[(\beta\Gamma_m\phi_m)\hat{\rho}] \pm \nonumber \\
&\pm& \int d^3r \Bigl[\frac{1}{2} ({\vec\nabla}\phi_m)^2 + U(\phi_m)\Bigr]\Bigr\}
\label{cedf}
\eea
with the nucleon mass $M$ and non-linear sigma-meson potentials $U(\phi_m)$ \cite{Lalazissis1997}. In Eq. (\ref{cedf}) the sign "+" corresponds to the scalar $\sigma$-meson, "-" to the vector $\omega$-meson, $\rho$-meson and photon, and the index "$m$" runs over the bosonic and Lorentz indices \cite{VretenarAfanasjevLalazissisEtAl2005}.
The variation of Eq. (\ref{Omega})
determines the operator of the nucleonic density with the 
eigenvalues 
of the Fermi-Dirac distribution:
\be
n_1(T) = n(\varepsilon_1, T) = \frac{1}{1 + \text{exp}\{\varepsilon_1 /T \}},
\label{FermiDirac}
\ee
where the number index runs over the complete set of the single-particle quantum numbers in the Dirac-Hartree basis including
the single-particle energies $\varepsilon_1 = {\tilde\varepsilon_1} - \mu$ measured from the chemical potential $\mu$.
%
In this work we consider non-superfluid nuclear systems, such as doubly-magic nuclei and nuclei at temperatures above the critical temperature when superfluidity vanishes. 

The small-amplitude particle-hole response function is described by the Bethe-Salpeter equation (BSE) \cite{Salpeter1951}:
\bea
{\cal R}(14,23) &=& {\cal G}(1,3){\cal G}(4,2) + \nonumber \\
&+& \sum\limits_{5678} {\cal G}(1,5){\cal G}(6,2)V(58,67){\cal R}(74,83)
\label{bse1}
\eea
adopted to the finite-temperature formalism. The number indices in Eq. (\ref{bse1}) include the single-particle variables and time: $1 = \{k_1,t_1\}$, and ${\cal G}(1,3)$ are the Matsubara temperature Green's functions of single particles defined for the imaginary time differences: $t_{13} = t_1-t_3$ ($0<t_{1,3}<1/T$) \cite{Abrikosov1965}. 
The interaction kernel $V(58,67)$ includes both the instantaneous and the time-dependent contributions, as in the zero-temperature case. In this work, the former is given by the meson-exchange interaction and the latter is represented, in the leading approximation, by the exchange of the correlated particle-hole pairs (phonons) between nucleons. Eq. (\ref{bse1}) can be rewritten as
\be
{\cal R}(14,23) = \tilde{\cal R}(14,23) + \sum\limits_{5678} \tilde{\cal R}(16,25){\cal W}(58,67){\cal R}(74,83),
\label{bse2}
\ee
in terms of the uncorrelated particle-hole propagator ${\tilde{\cal R}}(14,23) = {\tilde{\cal G}}(1,3){\tilde{\cal G}}(4,2)$
and the redefined interaction kernel ${\cal W}(14,23)$.
The uncorrelated particle-hole propagator ${\tilde{\cal R}}(14,23)$ is a product of two fermionic temperature mean-field Green's functions $\widetilde{\mathcal{G}}$
which, in the imaginary-time representation, read \cite{Abrikosov1965}:
\begin{eqnarray}
\label{Full Mean-Field}\widetilde{\mathcal{G}}(2,1)&=&\sum_{\sigma}\widetilde{\mathcal{G}}^{\sigma}(2,1),\\
\label{Component Mean-Field}\widetilde{\mathcal{G}}^{\sigma}(2,1)&=&-\sigma\delta_{{1}{2}}n(-\sigma\varepsilon_{{1}},T)
e^{-\varepsilon_{{1}}t_{21}}\theta(\sigma t_{21}),
\end{eqnarray}
where $t_{21}=t_{2}-t_{1}$ ($-1/T<t_{21}<1/T$), $\theta(t)$ is the Heaviside step-function and
the index $\sigma=+1(-1)$ denotes the retarded (advanced) component of $\widetilde{\mathcal{G}}$. 
The new interaction kernel decomposes as follows:
\bea
{\cal W}(14,23) = {\tilde V}(14,23) + V^e(14,23) +  \nonumber \\ 
+ {\tilde{\cal G}}^{-1}(1,3)\Sigma^e(4,2) +  \Sigma^e(1,3){\tilde{\cal G}}^{-1}(4,2),
\label{W-omega}
\eea
into the meson-exchange interaction $\tilde V$ specified below, the phonon-exchange term $V^e$ and the corresponding self-energy terms ${\tilde{\cal G}}^{-1}\Sigma^e$ and $\Sigma^e{\tilde{\cal G}}^{-1}$, such that
$V^e =\delta \Sigma^e / \delta{\tilde{\cal G}}$, in analogy to the BSE in the particle-hole channel at $T=0$ \cite{KamerdzhievTertychnyiTselyaev1997,Tselyaev2007, LitvinovaTselyaev2007,LitvinovaRingTselyaev2007}. At zero temperature, Eq. (\ref{bse2}) can be solved in the time blocking approximation \cite{Tselyaev1989,KamerdzhievTertychnyiTselyaev1997,Tselyaev2007} which reduces the Fourier transform of Eq. (\ref{bse2}) to a single frequency variable equation. The approximation is based on the time projection technique within the Green function formalism, which allows for decoupling of configurations of the lowest complexity beyond $1p1h$ (one-particle-one-hole), such as $1p1h\otimes phonon$ (particle-hole pair coupled to a phonon), from the higher-order ones.
However, the time projection operator introduced at $T=0$ \cite{Tselyaev1989} is not applicable for the finite-temperature case and, thus, has to be generalized. We found out in Refs. \cite{WibowoLitvinova2019, LitvinovaWibowo2018} that at $T>0$ the projection operator, 
which should be introduced into the integral part of Eq. (\ref{bse2}), has the following form:
\bea
{\Theta}(14,23;T) = \delta_{\sigma_1,-\sigma_2}\theta_{12}(T)\theta(\sigma_1t_{14})\theta(\sigma_1t_{23}),\nonumber \\
\theta_{12}(T) = n(\sigma_1\varepsilon_2,T)\theta(\sigma_1 t_{12}) + n(-\sigma_1\varepsilon_1,T)\theta(-\sigma_1 t_{12}),\nonumber \\
\eea
with $\sigma_k = +(-) 1$ for particle(hole) states and the extra $\theta_{12}(T)$ factor, as compared to $T=0$. 
Because of the diffuseness of the Fermi-Dirac distribution functions, this factor induces a soft blocking, which becomes sharp in the $T\to 0$ limit when $\theta_{12}(T) \to 1$.
After the 3-Fourier transformation, summing over the fermionic discrete energy variables and analytical continuation to the real-energy domain, the BSE for the proton-neutron response reads:

%
\begin{figure*}[ptb]
\begin{center}
\includegraphics[scale=0.55]{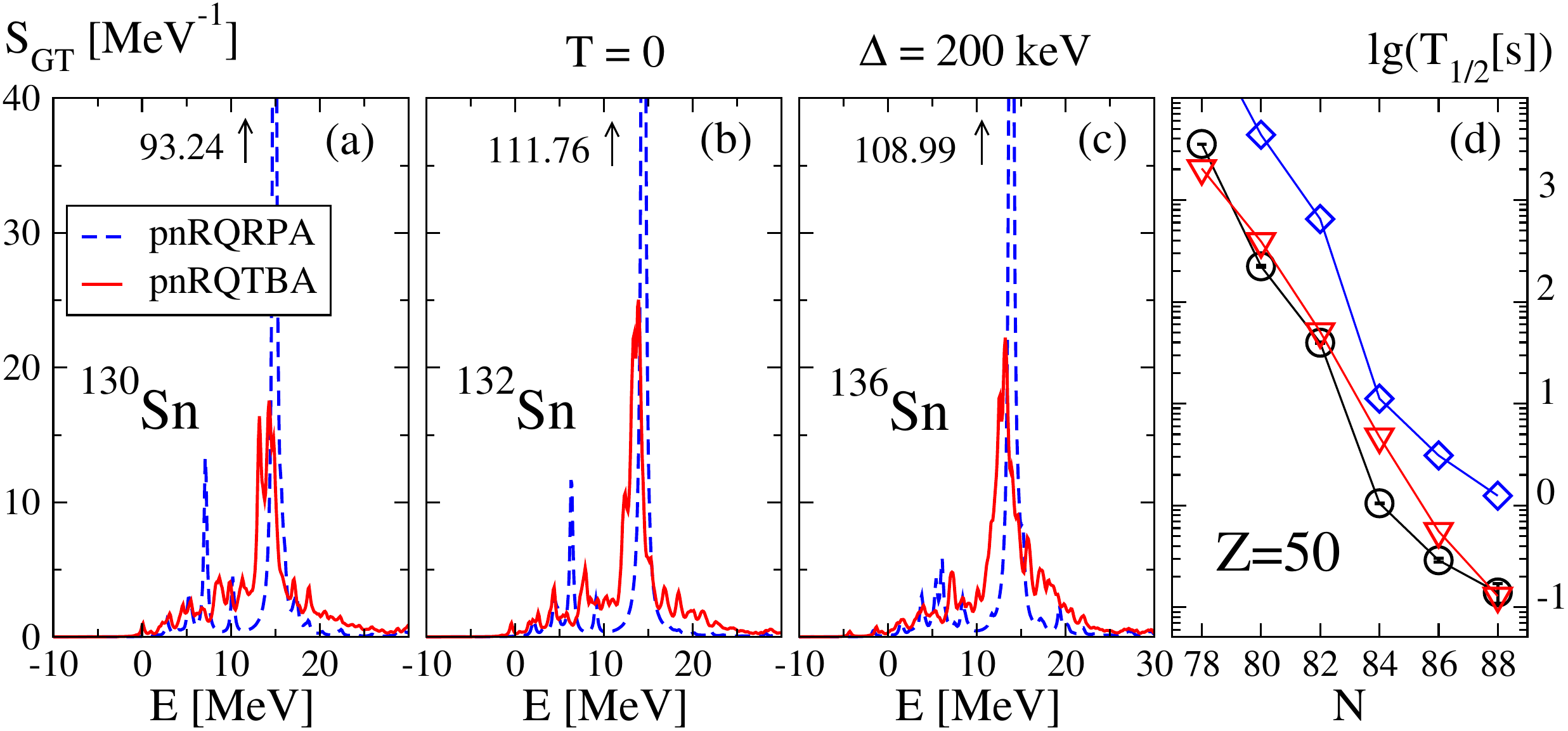}
\end{center}
\vspace{-0.5cm}
\caption{GT$_-$ strength distribution for $^{130,132,136}$Sn nuclei at zero temperature in the pnRQTBA, compared to the pnRQRPA (a-c). Beta decay half-lives in neutron-rich tin isotopes extracted from the pnRQRPA (diamonds) and pnRQTBA (triangles) strength distributions, compared to data (circles) \cite{nndc} (d).}  
\label{GTR_Sn}
\end{figure*}
%

\bea
{\cal R}_{pn',np'}(\omega,T) = \tilde{\cal R}_{pn}(\omega,T) \delta_{pp'} \delta_{nn'} + \nonumber\\
+ \tilde{\cal R}_{pn}(\omega,T) \sum\limits_{p^{\prime\prime}n^{\prime\prime}} 
 {\cal W}_{pn^{\prime\prime},np^{\prime\prime}}(\omega,T) 
{\cal R}_{p^{\prime\prime}n',n^{\prime\prime}p'}(\omega,T),\nonumber \\
\label{bse3}
\eea
where $\tilde{\cal R}(\omega,T)$ is the uncorrelated proton-neutron propagator
\begin{equation}
\tilde{\cal R}_{pn}(\omega,T) = \frac{n_{np}(T)}{\omega - \varepsilon_{p} + \varepsilon_{n} }, 
\label{resp0}
\end{equation}
$n_{np}(T) = n_{n}(T )- n_{p}(T)$ with the indices '$p$' and '$n$' of the proton and neutron states, respectively, and ${\cal W}(\omega,T)$ is the interaction amplitude: 
\be
{\cal W}_{pn',np'}(\omega,T) = \tilde{V}%
_{pn',np'}(T)
+ \Phi_{pn',np'}(\omega,T).
\label{W-omega-tba}%
\ee
In the charge-exchange channels the static part of the interaction $\tilde{V}$ is represented by the exchange of $\pi$ and $\rho$ mesons carrying isospin and the short-range Landau-Migdal term ${\tilde V}_{\delta\pi}$:
\be
{\tilde V} =  {\tilde V}_{\rho} + {\tilde V}_{\pi} + {\tilde V}_{\delta\pi} , 
\label{mexch}
\ee
where the the $\rho$-meson is parametrized according to Ref. \cite{Lalazissis1997}, the pion-exchange is treated as in a free space, and the strength of the last term is adjusted to the GTR in $^{208}$Pb \cite{PaarNiksicVretenarEtAl2004a}, in the absence of the explicit Fock term \cite{LongVanGiaiMeng2006,LiangVanGiaiMengEtAl2008,LiangZhaoRingEtAl2012}.
The PVC amplitude $\Phi(\omega,T)$ has the following form:
\begin{eqnarray}%
\Phi_{pn',np'}^{(ph)}(\omega,T)  = \frac{1}{n_{n'p'}(T) } 
\sum\limits_{p''n''\mu} \sum\limits_{\eta_{\mu}=\pm1}
\eta_{\mu} \xi^{\mu\eta_{\mu};p''n''}_{pn,p'n'}  
\nonumber \\
\times\frac{ \bigl(N(\eta_{\mu}\Omega_{\mu}) + n_{n''}(T)\bigr)\bigl(n(\varepsilon_{n''}-\eta_{\mu}\Omega_{\mu},T) - n_{p''}(T)\bigr)}{\omega-\varepsilon_{p''}+\varepsilon_{n''}-\eta_{\mu}\Omega_{\mu} }, 
\nonumber\\
\label{phiph}%
\end{eqnarray} 
with the phonon vertex matrices $\zeta^{\mu\eta_{\mu}}$ denoted as:
\be
\xi^{\mu\eta_{\mu};56}_{12,34} = \zeta^{\mu\eta_{\mu}}_{12,56}\zeta^{\mu\eta_{\mu}\ast}_{34,56}, \ \ \ \ \ \ 
\zeta^{\mu\eta_{\mu}}_{12,56} = \delta_{15}\gamma^{\eta_{\mu}}_{\mu;62} - \gamma^{\eta_{\mu}}_{\mu;15}\delta_{62},
\label{zeta} 
\ee
via the matrix elements of the particle-phonon coupling vertices, 
$\gamma_{\mu;13}^{\eta_{\mu}} = \delta_{\eta_{\mu},+1}\gamma_{\mu;13} + \delta_{\eta_{\mu},-1}\gamma_{\mu;31}^{\ast}$,
and the phonon frequencies $\Omega_{\mu}$. The 
index "$\mu$" includes the phonon quantum numbers, such as angular momentum, parity, and frequency. The vertices $\gamma_{\mu;13}$ and the frequencies $\Omega_{\mu}$ are extracted from the 
finite-temperature relativistic random phase approximation (FT-RRPA) as described in Refs. \cite{LitvinovaWibowo2018,WibowoLitvinova2019}. The bosonic occupation factors
$N(\Omega) = 1/ (e^{\Omega /T}-1)$ in Eq. (\ref{phiph}) are associated with the phonons emitted and absorbed in the intermediate states of the proton-neutron pair propagation.
The $(ph)$-component of the PVC amplitude (\ref{phiph}) includes the proton-neutron pairs constrained by the condition: $n_{pn}(T)\geq 0$, $n_{p'n'}(T)\geq 0$ while the $(hp)$-counterpart is calculated analogously \cite{WibowoLitvinova2019}.

{The spectral functions under study $S(\omega)$ related to the reduced transition probabilities $B_{\nu}$
\be
S (\omega) = -\frac{1}{\pi} \lim\limits_{\Delta \to 0} \text{Im} \Pi(\omega + i\Delta) = 
\sum\limits_{\nu} B_{\nu}\delta(\omega - \omega_{\nu}) 
\label{specfun}
\ee
are determined via the nuclear polarizability $\Pi(\omega)$ 
\be
\Pi (\omega + i\Delta) = \langle V^{(0)} {\cal R} V^{(0)\dagger}\rangle = 
\sum\limits_{\nu} \frac{B_{\nu}}{\omega - \omega_{\nu} + i\Delta} 
\label{polarizability}
\ee
by the Gamow-Teller (GT) and spin multipole (SL) external fields:
\bea
V^{(0)}_{GT_-} = \sum\limits_{i=1}^A \Sigma(i)\tau_-(i) \\
V^{(0)\lambda}_{SL\pm} = \sum\limits_{i=1}^{A} r^L(i)[\Sigma (i)\otimes Y_L(i)]^{\lambda}\tau_{\pm}(i),
\label{extfield}
\eea
 where $\Sigma$ is the relativistic spin operator.  
The final finite-temperature strength functions ${\tilde S}(\omega)$ contain an additional temperature correction \cite{WibowoLitvinova2019,Dzhioev2015}: 
\be
 \label{Strength}
\tilde{S}(\omega)=\frac{S(\omega)}{1-e^{-(\omega-\delta_{np})/T}},
\ee
where $\delta_{np} = \lambda_{np} + M_{np}, \lambda_{np}$  is the difference between
neutron and proton chemical potentials in the parent nucleus and $M_{np} = 1.293$ MeV is the neutron-proton mass
splitting. 
The functions $S(\omega)$ and ${\tilde S}(\omega)$ are formally singular. Therefore, for representation purposes the usual practice
 is to take a finite value of the imaginary part of the energy variable (smearing parameter) $\Delta$. It provides a smooth 
 envelope of the strength distribution and also averages over complex configurations which are not taken into account explicitly. The denominator
 in Eq. (\ref{Strength}) is important only for the excitation energies $|\omega-\delta_{np}| \leq T$ and is mostly close to unity for small temperatures under $\sim 2$ MeV. It is nearly negligible for the general  features of the strength distribution, however, it is taken into account in the calculations of the beta decay half-lives discussed below.
}
\begin{figure*}[ptb]
\begin{center}
\includegraphics[scale=0.55]{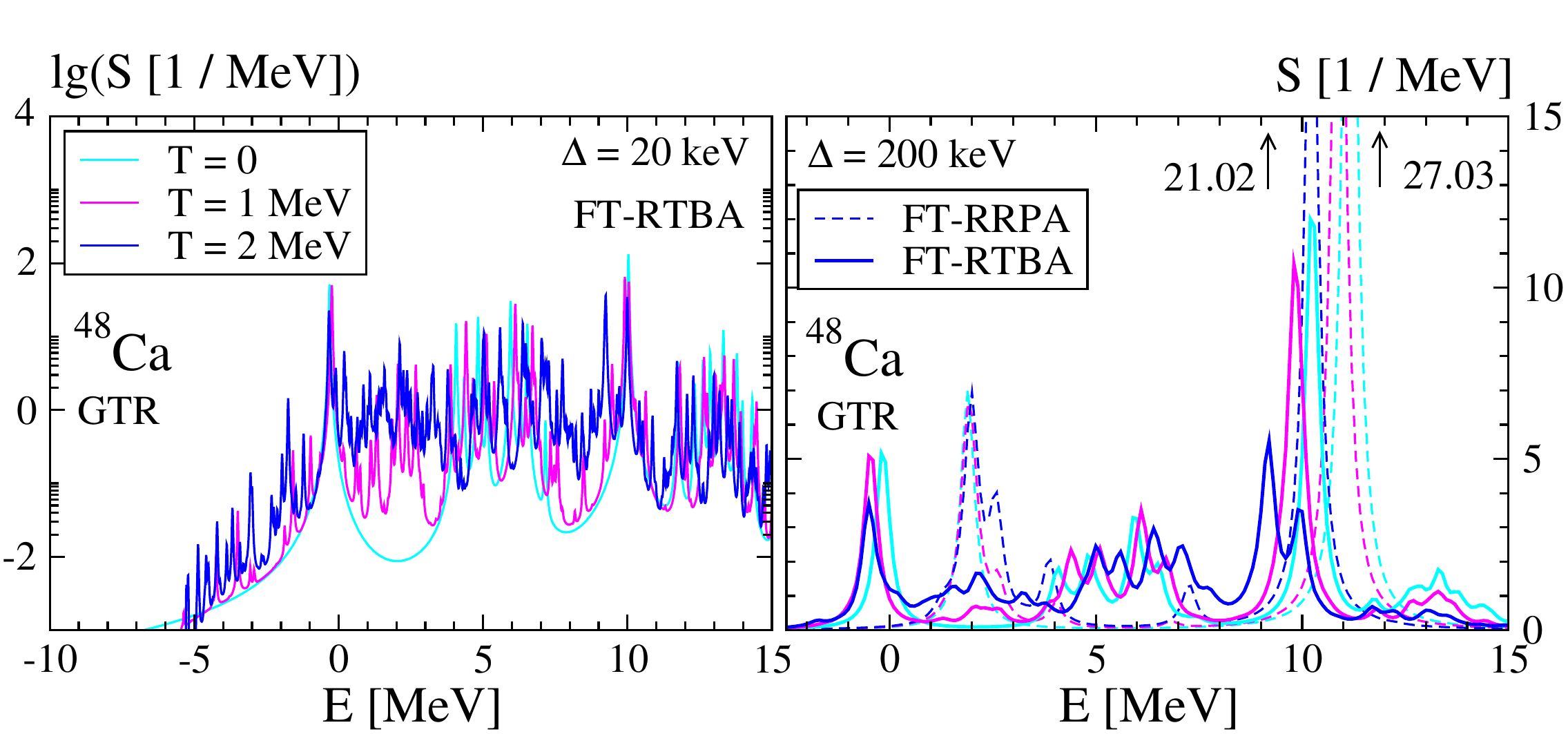}
\end{center}
\vspace{-0.5cm}
\caption{GT$_-$ strength distribution in $^{48}$Ca with respect to the ground state of the parent nucleus at various temperatures in the proton-neutron FT-RRPA (dashed curves) and FT-RTBA  (solid curves).}
\label{GTR_Ca}
\end{figure*}
\begin{figure*}[ptb]
\begin{center}
\includegraphics[scale=0.55]{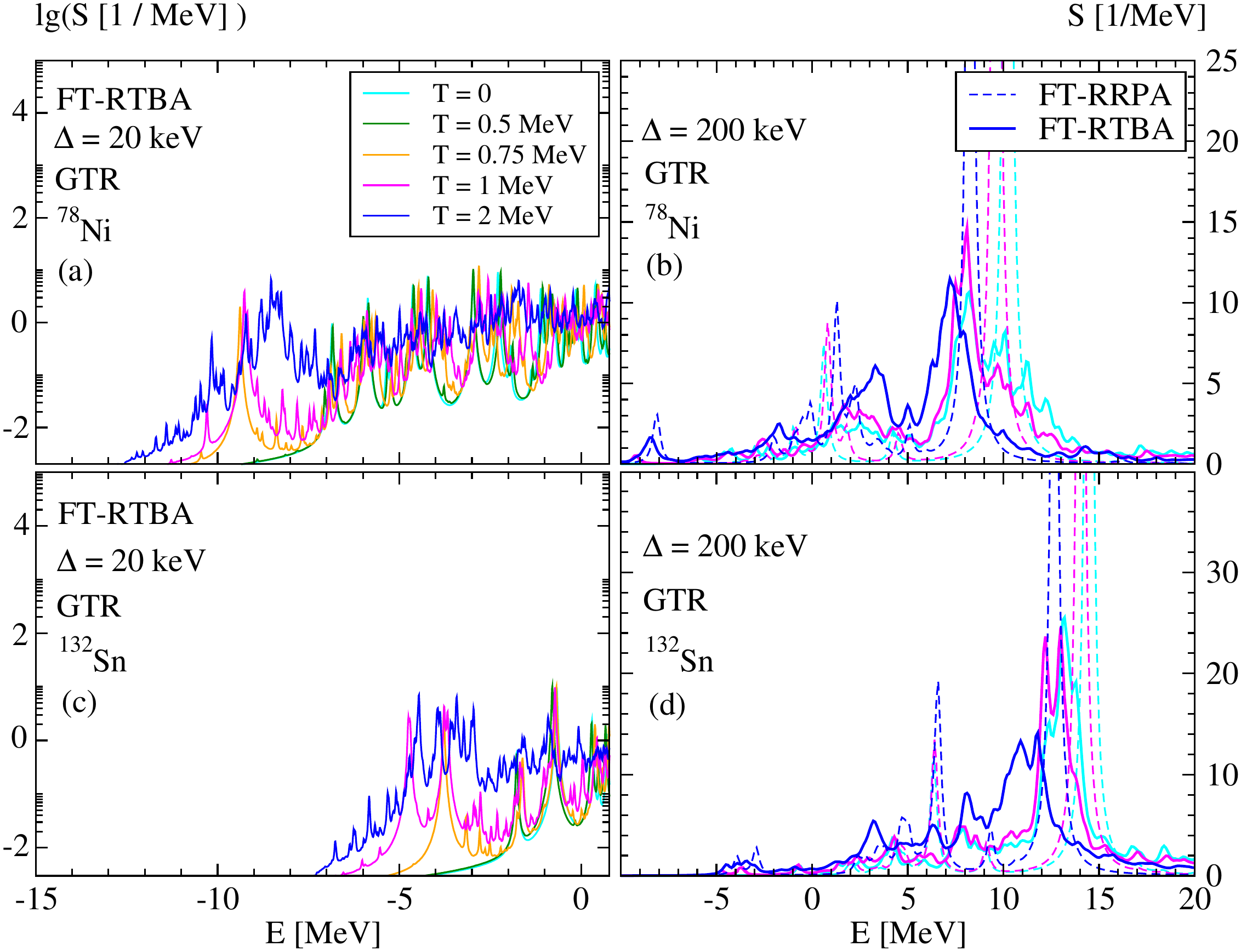}
\end{center}
\vspace{-0.3cm}
\caption{GT$_-$ strength distribution in $^{78}$Ni and $^{132}$Sn at various temperatures with respect to the ground states of the parent nuclei. See text for details.}
\label{GTR_NiSn}
\end{figure*}

{\it Results. \textemdash} The performance of the approach at $T = 0$ is illustrated in Fig. \ref{GTR_Sn} for the response of the semi-magic neutron-rich tin isotopes  to the GT$_-$ operator. 
 The details of 
these calculations are given in Ref. \cite{RobinLitvinova2016} and here and in the following the presented spectra are displayed on the excitation energy scales relative to the parent nuclei. One can see in  Fig. \ref{GTR_Sn}  that the PVC effects included in the proton-neutron relativistic quasiparticle time blocking approximation (pnRQTBA) 
produce a significant fragmentation of the GTR as compared to the proton-neutron relativistic QRPA (pnRQRPA). In turn,  this fragmentation redistributes the strength in the low-energy sector, in particular, in the $Q_{\beta}$ window and leads to faster beta decay, in agreement to experimental data \cite{nndc}. We will see in the following that at finite temperature the PVC term $\Phi(\omega,T)$ plays a similar role.

First calculations at $T>0$ within the proton-neutron finite-temperature relativistic time blocking approximation (FT-RTBA) (\ref{bse3}-\ref{phiph}) were performed for three closed-shell nuclei $^{48}$Ca, $^{78}$Ni, and $^{132}$Sn, for which we have obtained a very good description of data at $T=0$ \cite{RobinLitvinova2018}. In the latter work, the GT$_-$ strength functions for $^{48}$Ca and $^{132}$Sn were directly compared to data, together with the beta decay half-lives for $^{132}$Sn and also for $^{78}$Ni, where the GT strength distribution is still unavailable. In all cases, the PVC contributions were found crucial in reproducing experimental data. The agreement with data at $T=0$, together with the recent successful implementation of FT-RTBA for the neutral channel \cite{LitvinovaWibowo2018,WibowoLitvinova2019}, thus serves as a good benchmark for the present theory.

For the calculations at $T>0$ the same numerical scheme and model space truncations as in Refs. \cite{LitvinovaWibowo2018,WibowoLitvinova2019} were used in the present applications.
Fig. \ref{GTR_Ca} displays the GTR in a doubly-magic $^{48}$Ca at various temperatures. The right panel shows the general features of the GTR and its temperature evolution calculated within the proton-neutron FT-RRPA (dashed curves) and FT-RTBA (solid curves) with the imaginary part of the energy variable 
(smearing parameter) 
$\Delta$ = 200 keV. One can notice that the temperature increase induces an additional fragmentation of the overall strength distribution in both FT-RRPA and FT-RTBA as well as a shift of the entire distribution toward lower energies. The fragmentation occurs due to the thermal unblocking of transitions within the particle-particle and hole-hole pairs, which receive increasing numerators in Eq. (\ref{resp0}) with the temperature growth \cite{WibowoLitvinova2019}. As the calculations are self-consistent being based on the temperature-dependent mean field, due to the change of the single-particle energies with temperature the corresponding transition energies evolve accordingly as well as the proton and neutron chemical potentials. These effects contribute to the displacements of the entire GT distributions. Compared to FT-RRPA, the fragmentation effects due to the PVC in FT-RTBA  remain quite strong with the temperature increase for both high-energy and low-energy peaks. Calculations up to the temperature $T = 6$ MeV (not shown here) have revealed a continuation of these trends. The general features of the GTR obtained in the proton-neutron FT-RTBA calculations are consistent with the results of Refs. \cite{LacroixChomazAyik1998,LacroixChomazAyik2000} and with the model analyses of Refs. \cite{Sokolov1997, Stoyanov2004}. The left panel of Fig. \ref{GTR_Ca} displays a detailed fine structure of the GT$_-$ strength, which was obtained in FT-RTBA calculations with $\Delta$ = 20 keV, on the logarithmic scale. Here one can observe clearly a large amount of the new states emerging with more and more of the thermal unblocking in both high and low-energy sectors. 

\begin{figure*}[ptb]
\begin{center}
\includegraphics[scale=0.55]{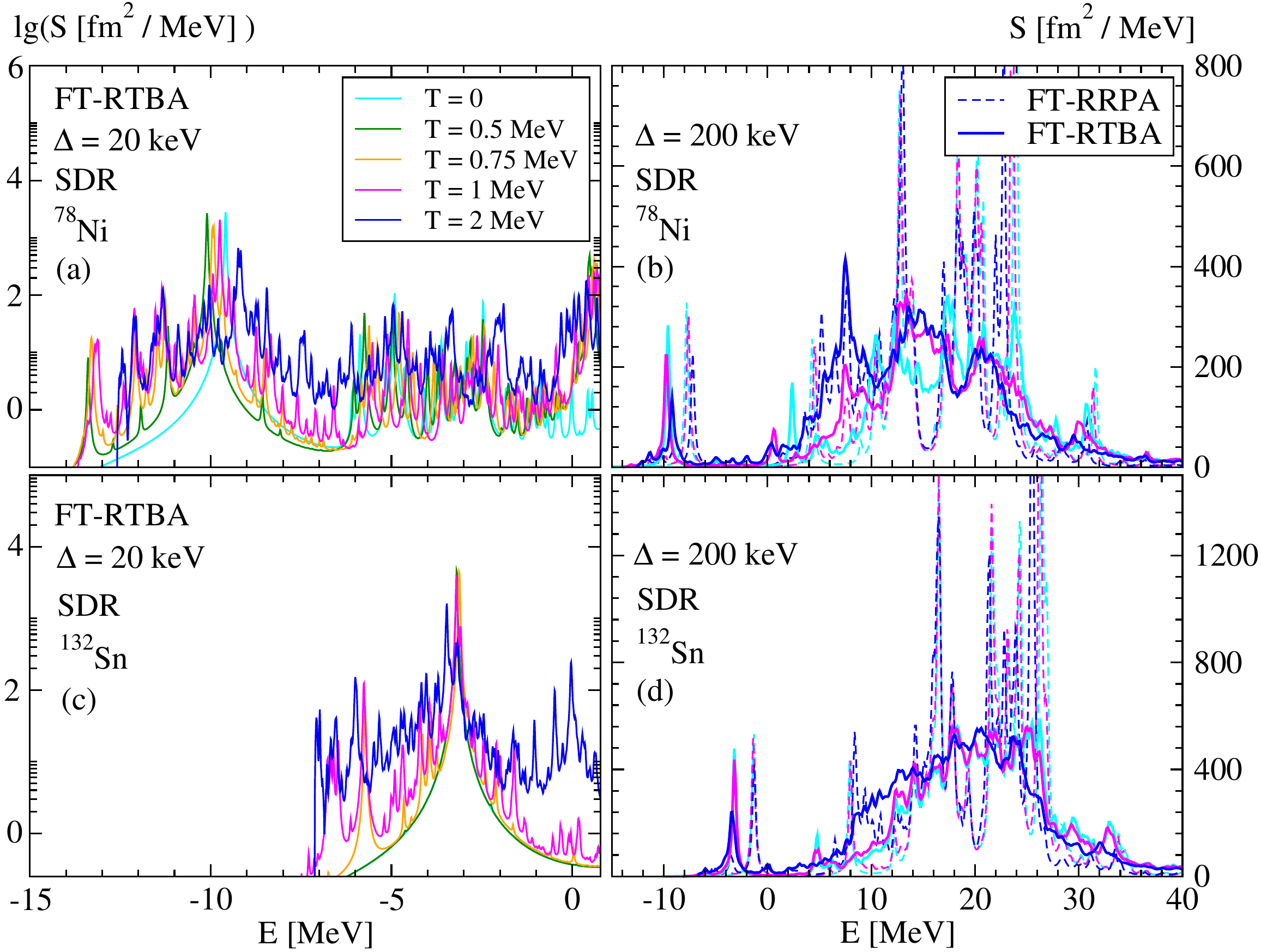}
\end{center}
\vspace{-0.3cm}
\caption{
Same as in Fig. \ref{GTR_NiSn}, but for the spin dipole resonance (SDR).
}
\label{SDR_NiSn}
\end{figure*}
%

In Fig. \ref{GTR_NiSn} the calculated GT$_-$ response is shown for $^{78}$Ni and $^{132}$Sn nuclei. The right panels (b, d) display the GT$_-$ strength distributions calculated with $\Delta$ = 200 keV 
in the proton-neutron FT-RRPA and FT-RTBA. Similarly to the case of $^{48}$Ca, the thermal and the PVC effects are clearly visible and cause general fragmentation of the GTR as well as its spread toward low energies. The left panels (a, c) demonstrate the temperature evolution of the fine structure of the GTR in these nuclei by showing the proton-neutron FT-RTBA calculations with $\Delta$ = 20 keV on a smaller temperature grid within the $Q_{\beta}$ window. The enhancement of the GT$_-$ strength in this energy region with the temperature growth is signaling about the increasing beta instability of both nuclei.  Presenting the GT$_-$ strength distributions on a finer temperature grid in panels (a,c) of Fig. \ref{GTR_NiSn} allows one to see, for instance, at which temperatures the thermal unblocking becomes strong enough to induce the formation of new states in the $Q_{\beta}$ window and, thus, should start to influence the beta decay rates. In the cases of  $^{78}$Ni and $^{132}$Sn nuclei the low-energy GT$_-$ strength distributions change notably between $T =$ 0.5 MeV and $T =$ 0.75 MeV, thus, the beta instability is expected to increase starting from these temperatures. This points out that in modeling astrophysical processes, which occur at these and higher temperatures, one has to take into account the temperature dependence of the nuclear GT$_-$ transitions, in addition to the thermal effects of the environment.

{The GT$_-$ strength distributions calculated with the small value of the smearing parameter and displayed in the left panels of Fig. \ref{GTR_NiSn} allow an extraction of the $B_{GT}$ values that can be compared to the available experimental data. As follows from the relations (\ref{specfun},\ref{polarizability}), $B_{GT}(\omega_{\nu}) \equiv B_{\nu} \approx S(\omega_{\nu})\pi\Delta$, i.e. the reduced transition probabilities are defined by the peak values of $S(\omega)$ multiplied by the smearing parameter $\Delta$ for relatively small values of $\Delta$. In this case $S(\omega)$ scales linearly with $\Delta$, so that their product remains constant and independent on $\Delta$. For example, at $T=0$ for the three lowest states in $^{132}$Sn we have the $B_{GT}$ values of $B_{GT}^{(1)}$=0.037, $B_{GT}^{(2)}$=0.607 and $B_{GT}^{(3)}$=0.147 units. The energies of these states with respect to the RMF ground state of $^{132}$Sb are $E^{(1)}$ = 0.01 MeV, $E^{(2)}$ = 1.11 MeV and $E^{(3)}$ = 2.08 MeV.  The latter two  excitations may be compared to the experimentally observed ones at $E^{(1)}$ = 1.325 MeV with $B_{GT}^{(1)}$=0.364 and  $E^{(2)}$ = 2.268 MeV with $B_{GT}^{(2)}$=0.0577 \cite{nndc}. The higher $B_{GT}$ values and the presence of the lowest relatively weak GT$_-$ state obtained in RTBA may be artifacts of the incomplete theoretical description. We already know that coupling to the charge-exchange phonons \cite{RobinLitvinova2018}, complex ground state correlations \cite{Robin2019} and higher-order configurations \cite{LitvinovaSchuck2019} cause additional redistribution, upward shift and further fragmentation of the strength. 
We also note that in the RRPA, which does not include the PVC, the beta decay of $^{132}$Sn is very strongly hindered (see Fig. \ref{hlives} below)  because this approach produces only one weak state just below the upper integration bound of the Q$_{\beta}$ window. In this context, the RTBA result demonstrates a significant improvement. Since the finite-temperature generalization of the RTBA implies a sophisticated derivation in terms of the Matsubara Green functions formalism as well as a rather complicated numerical implementation, as the first step in that direction we generalized only the simplest version of the RTBA (the so-called resonant RTBA with neutral phonons) for finite temperatures. That is why we do not discuss more sophisticated models at $T=0$ here, however, they can be potentially generalized to the case of finite temperature in a future work.}
%

The GT$_-$ transitions are not the only ones which contribute to the beta decay. Indeed, as it was shown in a number of works, the 
first-forbidden transitions also play a role in this process \cite{Behrens1971,Warburton1994,Zhi2013,Marketin2016a}. In order to evaluate their contribution, we have calculated the response of $^{78}$Ni and $^{132}$Sn nuclei
to the spin dipole (SD) operator $V^{(0)}_{SD1_-} = \sum_{i=1}^{A} r(i)[\Sigma(i)\otimes Y_1(i)]^{J}\tau_{-}(i)$ for $J^{\pi} = 0^-, 1^-, 2^-$. The corresponding strength functions calculated with and without nuclear charge form factor reflecting the influence of the proton structure, together with the isovector dipole response and Dirac $\gamma_5$ matrix elements, define the contribution of the FF transitions to the beta decay rates \cite{Behrens1971}. Fig. \ref{SDR_NiSn} shows the spin dipole response (SDR) summed over the angular momenta $J = 0, 1, 2$ as an example of the typical behavior of the FF transitions. Like in the previous figure, the right panels (b, d) display the overall SDR up to high excitation energy and the left panels (a, c) emphasize the fine structure of the SDR in the $Q_{\beta}$ window on a finer temperature grid. We see that the temperature increase broadens the overall SDR distribution and slightly shifts the entire spectrum toward lower energies. Compared to the FT-RRPA calculations, the strength is also strongly fragmented at all temperatures. The fine structure of the low-energy part of the SDR given in panels (a, c) shows a remarkable sensitivity of the FF transitions to temperature, especially in the case of $^{78}$Ni where they change noticeably already at $T=0.5$ MeV. More strength appears in the $Q_{\beta}$ window with the temperature growth while at $T=2$ MeV the lowest states are visibly pushed up in energy. The latter occurs due to the decrease of the difference between the proton and neutron chemical potentials. 


%
\begin{figure}[ptb]
\vspace{0.5cm}
\begin{center}
\includegraphics[scale=0.37]{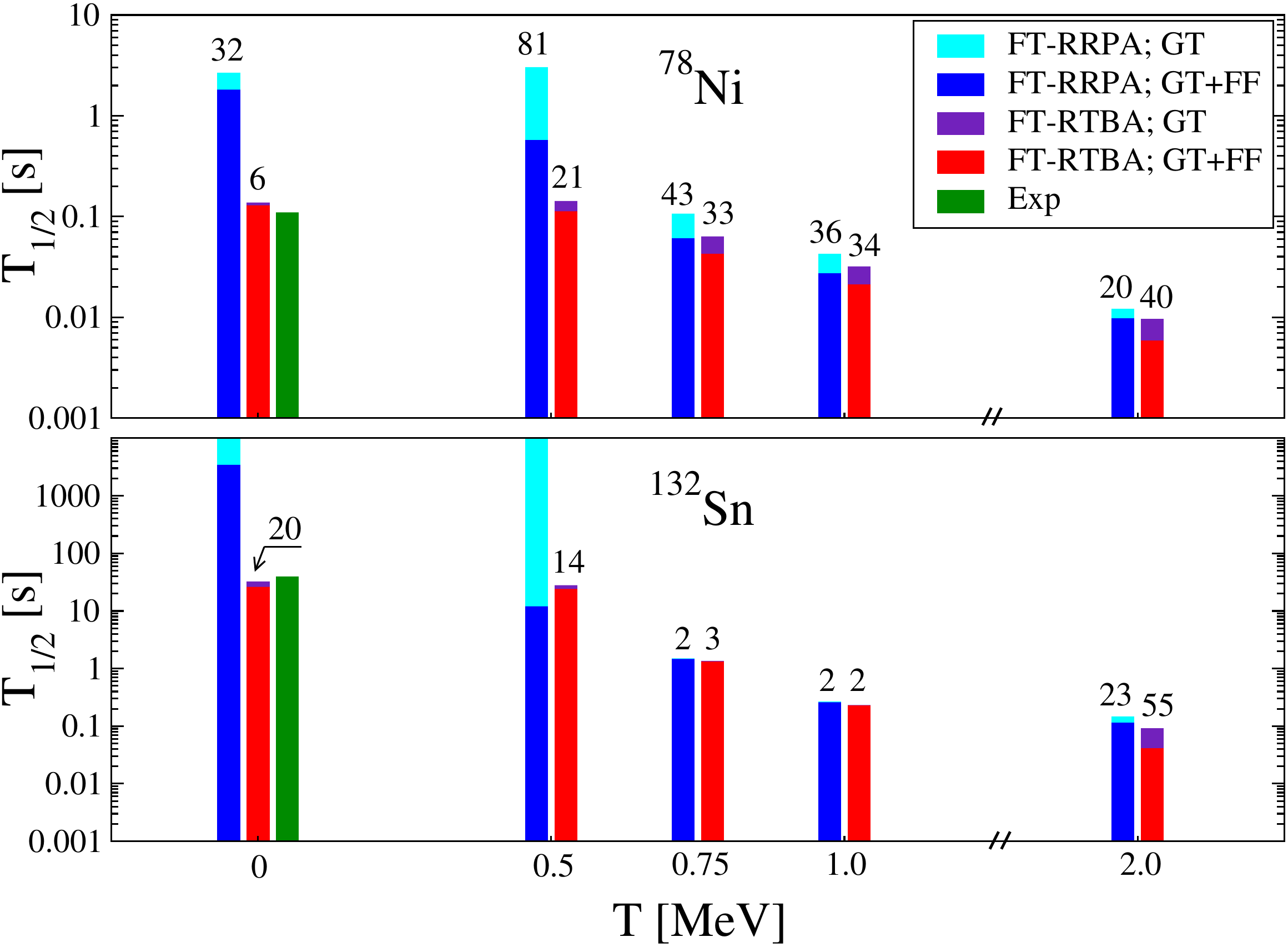}
\end{center}
\vspace{-0.5cm}
\caption{Beta decay half lives of $^{132}$Sn and $^{78}$Ni at various temperatures for electron density $\text{lg}(\rho Y_e)$ = 7.}
\label{hlives}
\end{figure}
\begin{figure*}[ptb]
\vspace{0.5cm}
\begin{center}
\includegraphics[scale=0.55]{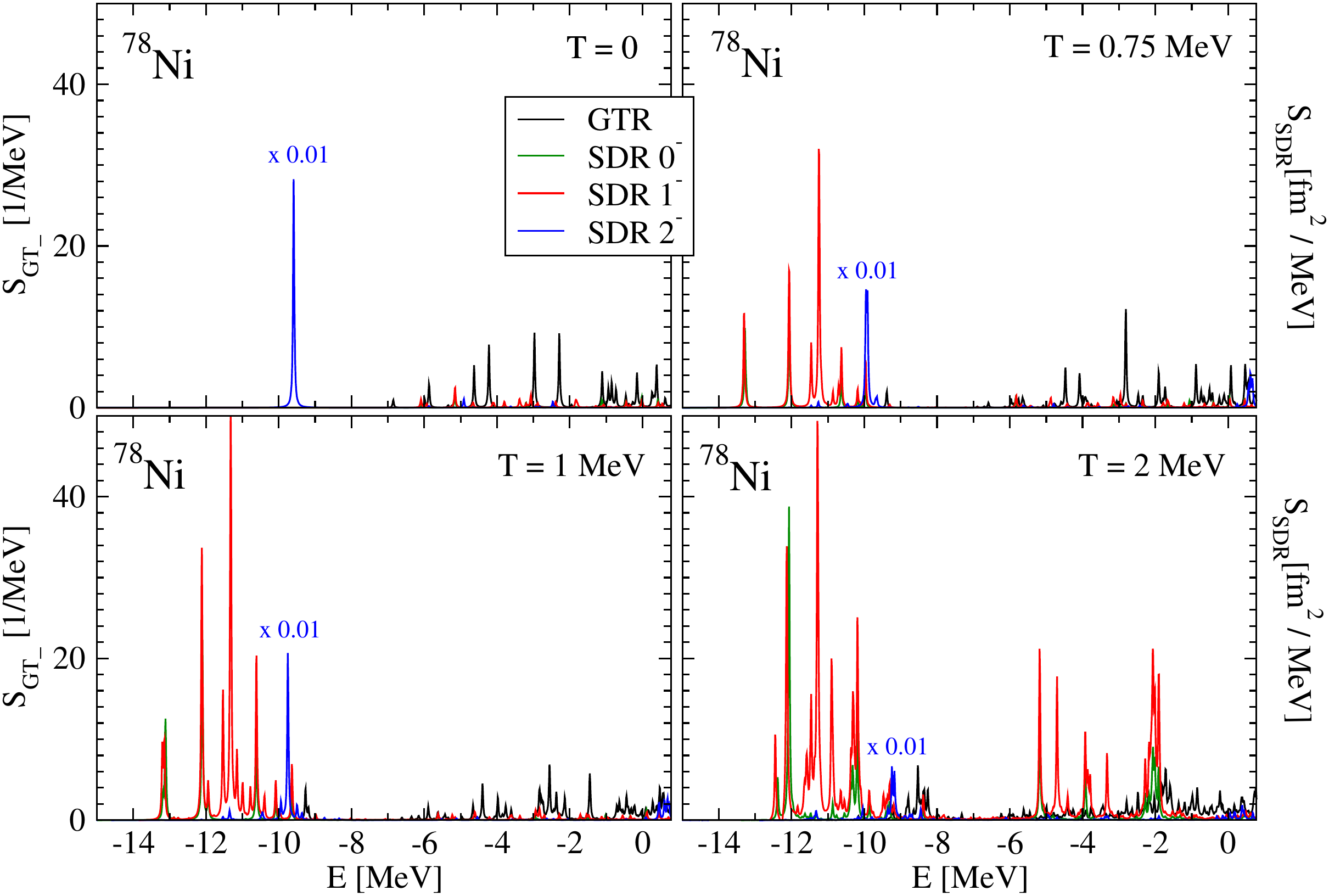}
\end{center}
\vspace{-0.5cm}
\caption{The $0^-,1^-$, and $2^-$ components of the spin dipole response in comparison with the GT$_-$ response in $^{78}$Ni at various temperatures. The spectral functions are computed with 
$\Delta$ = 20 keV to better illustrate and compare the effects of the thermal unblocking mechanism in different channels.}
\label{gtr_sdr}
\end{figure*}

The nuclei $^{78}$Ni and $^{132}$Sn play a very important role of the so-called waiting points in the r-process nucleosynthesis while $^{78}$Ni is also relevant to the pre-collapse phase of the core-collapse supernovae (CCSN). Ultimately, identical simulation frameworks are to be used for both the neutron star mergers, where the r-process occurs, and the CCSN \cite{Aprahamian2018}.
Thus, to illustrate the temperature dependence of the beta decay rates, we adopt some fixed medium values of electron density $\rho$ and electron-to-baryon ratio $Y_e$, such as $\text{lg}(\rho Y_e)$ = 7, from the Fuller, Fowler and Newman (FFN) temperature-density grid \cite{Fuller1982,Fuller1985,Langanke2000}.
The results for $^{132}$Sn and $^{78}$Ni are summarized in Fig. \ref{hlives}. 
The values of $T_{1/2}$ for $^{78}$Ni and $^{132}$Sn are associated with the heights of the histograms in the upper and lower panels, respectively.  
For each temperature, we show the beta decay half-lives extracted from the  proton-neutron FT-RRPA and FT-RTBA strength distributions with the smearing parameter $\Delta = $ 20 keV, which ensures converged values of the half-lives. The half-lives are given with (GT+FF) and without (GT) contributions of the FF transitions. The latter contributions are indicated in Fig. \ref{hlives} in percentage with respect to the total beta decay rates $\lambda$ defined as $\lambda = \text{ln}2/T_{1/2}$. These ratios are not given only for (FT)-RRPA in $^{132}$Sn at $T=0$ and $T=0.5$ MeV because of practically absent GT$_-$ transitions. The half-lives have been evaluated according to 
Refs. \cite{Behrens1971,Zhi2013,Marketin2016a} while accounting for the temperature dependence of the leptonic phase space and detailed balance as in Refs. \cite{Langanke2001,Dzhioev2015}. However, in contrast to the Refs. \cite{Zhi2013,Marketin2016a}, in our calculations no quenching factors were used for the transition matrix elements or for the axial vector coupling constant, and no adjustable proton-neutron pairing was introduced. As in Ref. \cite{RobinLitvinova2018}, at $T =$ 0 the proton-neutron RRPA strongly overestimates the $T_{1/2}$ values in both nuclei, in particular, $^{132}$Sn looks almost stable, however, the proton-neutron RTBA brings them in a very good agreement with experimental observations. In the present calculations we found that the inclusion of the first-forbidden transitions shortens the half-lives slightly further, 
but the inclusion of ground state correlations associated with the PVC ($GSC_{PVC}$) should correct for this small shortcoming \cite{Robin2019}. 
{We also note that the relative contributions of the FF transitions to the beta decay rates of $^{78}$Ni and $^{132}$Sn at  $T=0$ are consistent with the trends discussed in Ref. \cite{Marketin2016a}, in particular, with the shell-model calculations of Ref. \cite{Zhi2013}. 
}
At $T>0$ one can observe a gradual decrease of $T_{1/2}$ in both nuclei after $T = 0.5$ MeV. A small increase of the total half-lives at this temperature occurs because of the electron Fermi-Dirac distribution factor \cite{Langanke2001} while the nuclear spectra change relatively little. However, at higher temperature the enhancement of the GT$_-$ and FF transitions seen in Figs. \ref{GTR_NiSn},\ref{SDR_NiSn} at low energies starts to be important. The contribution of FF transitions is also increasing gradually after $T>0.5$ MeV in $^{78}$Ni while in $^{132}$Sn it has a minimum at $T= 1$ MeV because of strong mutual cancellation of the associated matrix elements. When going from $T = 0$ to $T = 2$ MeV, the overall FT-RTBA half-lives decrease by a factor of 22 and 632 in $^{78}$Ni and $^{132}$Sn, respectively, while the FF transitions contribute to the beta decay rates by 40\% and 55\%  at $T=2$ MeV, compared to 6\% and 20\% at $T=0$. Open-shell nuclei are expected to be even more sensitive to low temperatures, therefore, future developments should address effects of superfluid pairing.

{In Fig. \ref{gtr_sdr} we show separately the $0^-,1^-$, and $2^-$ components of the spin dipole resonance together with the the GT$_-$ response in $^{78}$Ni at various temperatures in the Q$_{\beta}$ window. Since the SD operator contains a radial form factor "r", the SD strength has different units, that is reflected on the plot (the 2$^-$ strength is considerably quenched to make the comparison possible).  One can see that at $T=0$ there is a $2^-$ state at low energy which brings the dominant among the FF transitions contribution to the beta decay rates, which are discussed above. With the temperature increase this $2^-$ state undergoes fragmentation, while new states of the $0^-$ and $1^-$ character appear in the Q$_{\beta}$ window due to the thermal unblocking. At the same time, the GT$_-$ strength shows a similar growth and redistribution in the low-energy domain.  The remaining six FF strength functions contributing to the beta decay \cite{Behrens1971,Zhi2013,Marketin2016a} are variations of the three SDR components with different radial form factors or the absence of the spin-flip, i.e. demonstrate a similar behaviour.  The calculation of the T$_{1/2}$ contains all these contributions with different coefficients, and the cumulative growth of contributions of all the FF strength functions in the Q$_{\beta}$ window may be faster than the one of the GT$_-$ strength, although the latter remains dominant in the considered temperature range. This can explain the increasing role of the FF transitions with temperature. 
}

%
%

{In this work we  present the FT-RTBA calculations for the strength functions only in the $\beta^-$ branch. They are of a great astrophysical interest as a key ingredient for the r-process nucleosynthesis, however, the typical temperatures for it are of the order of 100 keV. The nuclear structure impact of such temperatures is relatively small.  Similar calculations for the $\beta^+$ decay and electron capture would possibly have a stronger astrophysical impact as they occur, for instance, in the core-collapse supernovae within the temperature range of 0-2 MeV. However, as we discuss in Ref. \cite{Robin2019}, for nuclei with a neutron excess the $\beta^+$ branch may require a more sophisticated approach than the resonant FT-RTBA of the present form. Such an extended approach should include at least complex ground state correlations caused by the PVC effects, which are found to be  essential for the description of the $\beta^+$ processes in nuclei with a neutron excess. The finite-temperature generalisation of the FT-RTBA extended by the $GSC_{PVC}$ is currently not existing, but can be developed in the future as the next step after our present advancement. The continuum effects \cite{Kamerdzhiev1998,LitvinovaTselyaev2007} and configurations higher than ph$\otimes$phonon included in the conventional RTBA should also play a role in the description of both $\beta^-$ and $\beta^+$ branches, and they will be considered in the future work as well.
}

{\it Summary. \textemdash} The nuclear charge-exchange finite-temperature response theory 
is advanced beyond the one-loop approximation. 
The new approach is designed for computing the nuclear proton-neutron response at finite temperature taking into account the PVC spreading mechanism, in addition to the Landau damping. The time blocking technique, which was generalized lately to the case of finite temperature in Refs. \cite {LitvinovaWibowo2018,WibowoLitvinova2019} and now adopted to the 
 isospin-transfer excitations, allows for a numerically stable and executable calculation scheme, which is implemented on the base of 
quantum hadrodynamics in a parameter-free framework. 

 The temperature evolution of the spin-isospin response
in closed-shell nuclei  $^{48}$Ca, $^{78}$Ni, and $^{132}$Sn is investigated quantitatively and discussed in detail for the temperature range between zero and 2 MeV, which is relevant for astrophysical modeling. A remarkable enhancement of the Gamow-Teller and spin dipole transitions at lowest excitation energies is found already at moderate temperatures, while the fragmentation effects due to the PVC mechanism remain strong. We show that this enhancement, as a consequence of the thermal unblocking, gives rise to the shortening of the beta decay half-lives with the temperature increase in hot stellar environments and to the possibly increasing importance of the first forbidden transitions with temperature. 
Being well constrained at $T=0$ and benchmarked in neutral-channel calculations at $T>0$, the developed approach can provide an accurate description of the proton-neutron response, beta decay and electron capture rates in a wide range of temperatures and densities. Thus,  it can support modeling of various astrophysical objects, from supernovae to neutron star mergers.


The authors greatly appreciate discussions with A. Dzhioev, G. Mart{\'i}nez-Pinedo, R. Surman and M. Famiano. This work is partly supported by US-NSF Grant PHY-1404343, NSF Career Grant PHY-1654379, by the Institute for Nuclear Theory under US-DOE Grant DE-FG02-00ER41132 and by JINA-CEE under US-NSF Grant PHY-1430152.
%

\bibliography{Bibliography_Jul2018}
\end{document}